\documentclass[twocolumn,floatfix, showpacs]{revtex4}

\usepackage[final]{epsfig}
\usepackage{amsmath}
\begin{document}
 
\title{Hard dihadron correlations in heavy-ion collisions at RHIC and LHC}
 
\author{Thorsten Renk}
\email{thorsten.i.renk@jyu.fi}
\author{Kari J.~Eskola}
\email{kari.eskola@phys.jyu.fi}
\affiliation{Department of Physics, P.O. Box 35, FI-40014 University of Jyv\"askyl\"a, Finland}
\affiliation{Helsinki Institute of Physics, P.O. Box 64, FI-00014 University of Helsinki, Finland}


\begin{abstract}
High transverse momentum ($P_T$) processes are considered to be an important tool to probe and understand the medium produced in ultrarelativistic heavy-ion collisions via the interaction of hard, perturbatively produced partons with the medium.
In this context, triggered hard dihadron correlations constitute a class of observables set between hard single inclusive hadrons (dominated by the leading jet fragments) and fully reconstructed jets --- while they probe some features of the perturbative QCD evolution of a parton shower in the medium, they do not suffer from the problem of finding a suitable separation between soft hadrons coming from perturbative jets and soft hadrons coming from the non-perturbative medium as the identification of full jets does. On the other hand, the trigger requirement introduces non-trivial complications to the process, which makes the medium-modification of the correlation pattern difficult and non-intuitive to understand. In this work, we review the basic physics underlying triggered dihadron correlations and make a systematic comparison of several combinations of medium evolution and parton-medium interaction models with the available data from 200 AGeV Au-Au collisions at RHIC. We also discuss the expected results for 2.76 ATeV Pb-Pb collisions at the LHC. 
\end{abstract}
 
\maketitle

\section{Introduction}

High transverse momentum ($P_T$) hadron production in the context of ultrarelativistic heavy-ion (A-A) collisions is considered to constitute an important set of observables to probe properties of the medium produced in such collisions. Often the focus is on the nuclear suppression 
of hard hadrons in A-A collisions compared with the scaled expectation 
from p-p collisions, which is due to loss of energy from the hard parton by 
interactions with the soft medium (see e.g. \cite{Tomo1,Tomo2,Tomo3,Tomo4,Tomo5}). The relevant observable is the nuclear 
suppression factor $R_{AA}$ which, as well as its dependence on the event plane in noncentral collisions, has been quite thoroughly investigated in the context of various models \cite{SysJet1,SysJet2,Dihadron,SysHyd2,JetHydroSys,SpectraLHC}.

Another (visually more stunning) manifestation of the medium modification of high $P_T$ partons is the appearance of hadronic monojets in the medium if one parton out of a primary hard parton pair is quenched due to interactions with the medium. While such monojets have now been observed directly by the ATLAS and CMS collaboration in 2.76 ATeV Pb-Pb collisions at the LHC \cite{ATLAS,CMS}, the phenomenon was also observed via triggered hard dihadron correlations by the STAR experiment at RHIC \cite{STAR-dihadron1} and more differentially measured in \cite{STAR-DzT}.

From a theoretical perspective, the treatment of hard dihadron correlations is lagging behind hard single inclusive hadron production, as the correlation observable requires a substantially more involved numerical treatment due to the emergence of various biases which need to be taken into account consistently. First results \cite{Dihadron_orig,Dihadron,Dihadron-NLO} seemed to indicate that models which describe single hadron suppression well also succeed in the description of dihadron correlations, i.e. the results implied that there would not be a substantial new property of the medium probed in dihadron correlations. However, it was later realized that dihadron correlations provide tighter constraints on model parameters than single hadron suppression \cite{Constraints} and that they are more sensitive to the pathlength dependence of the parton-medium interaction \cite{Elastic}. The dependence of the correlation on the angle of the back-to-back event with the bulk event has also been studied in \cite{Dihadron_ang}, however no clear picture in comparison with data has emerged so far.

It is the aim of this work to provide an updated view on the potential of hard dihadron correlations as an observable to probe the interaction of hard partons with an evolving bulk medium. This is done in the light of both the ongoing LHC heavy ion program with a substantially extended reach in $\sqrt{s}$ as compared to the 200 AGeV available at RHIC and the arrival of in-medium shower evolution Monte-Carlo (MC) codes \cite{JEWEL,YaJEM1,YaJEM2,QPYTHIA,MARTINI}. Such codes are better suited for the treatment of the contribution of subleading jet hadrons to the correlation strength than the models based on leading parton energy loss which have often been used for the purpose so far.

\section{The physics of hard dihadron correlations}

Triggered hard dihadron correlations are interesting observables because they reflect an important part of the Quantum Chromodynamics (QCD) dynamics of a hard scattering process. In leading order (LO) perturbative QCD (pQCD), a hard process results in a back-to-back pair of highly virtual partons. Subsequently, a partonic shower with decreasing virtuality scale develops for each of the two parent partons until a non-perturbative hadronization scale is reached and the parton showers turn into jets of hadrons. The total momentum balance of these two jets contains information about the primary hard QCD process whereas the momentum distribution of hadrons inside the jets contains information about the QCD dynamics of partonic shower and hadronization.

Triggering on a hard hadron in a given momentum range thus on average picks the most energetic hadron of the two back-to-back jets. Correlations with subleading hadrons on the near side (i.e. in the trigger hemisphere) allow to probe into the QCD evolution of a jet, whereas correlations of the trigger hadron with the leading and subleading hadrons on the away side (i.e. opposite to the trigger hemisphere) allow to probe in addition the primary scattering process.

The role of the trigger is twofold: First, it provides some amount of information about the kinematics at the hard process. In the case of $\gamma$-h or Z-h correlations where the QCD evolution of the trigger side is absent, the trigger momentum sets rather stringent constraints for the away side initial parton kinematics event by event. In the case of hadron-hadron correlations the constrains are probabilistic and the trigger momentum can only constrain the parton kinematics on average. Second, selecting only events in which a trigger is seen introduces a bias --- the subclass of back-to-back hadron jets in which a hadron falls into a certain momentum range is different from the class of all back-to-back jet events. This bias can be systematically exploited to change the observed structure of the events.

When the hard process is embedded into a hot and dense medium such as created in ultrarelativistic A-A collisions, rather general uncertainty relation arguments indicate that the primary hard scattering process cannot be substantially changed by the medium, i.e. the corresponding rate is expected to be perturbatively calculable and can hence serve as a 'standard candle'. However, the subsequent evolution of a parton shower probes time and distance scales characteristic for the medium evolution. The shower is expected to be modified by the medium and consequently to carry information about both properties of the medium and the nature of parton-medium interaction. Finally, for light hadrons  or sufficiently hard momentum scales, the hadronization process can be expected to take place outside the medium. 

There is no information in hard dihadron correlation that cannot in principle be gained by studying full back-to-back jets. However, in practice in unbiased jets there is significant flow of energy and momentum through low $P_T$ hadrons for which, in a medium, problems such as distinguishing jet hadrons from medium hadrons or the possibility of in-medium hadronization (which due to its non-perturbative nature is poorly understood) occur. Hard dihadron correlations on the other hand allow to probe part of the QCD dynamics of jets by focusing on the leading jet fragments which can be required to be in a momentum range where such complications are absent.

\subsection{Observables and terminology}

The direct observables in dihadron correlation measurements are conditional yields, the so-called yield per trigger (YPT) on the near and away side. These depend crucially on the momentum range required for a trigger hadron. The YPT do not directly reflect any suppression of the observed rate of triggered events in A-A collisions compared to a p-p reference (which is closely related to the nuclear suppression factor $R_{AA}$). Thus, in an initial state suppression picture where a back-to-back parton event is either suppressed or never feels a medium, $R_{AA}$ can be at an arbitrarily low value while the  YPT is unchanged in the medium.

The near and away side YPT is often binned in momentum windows, but sometimes also in terms of the fraction of the associate hadron momentum divided by trigger hadron momentum, $z_T = P_T^{assoc}/P_T^{trig}$. Since $z_T$ has a probabilistic connection to the fractional momentum $z = P_T^{had}/p_T^{part}$ of a hadron produced from a parton with momentum $p_T^{part}$, the away side distribution binned in terms of $z_T$ is often called $D(z_T)$ similar to the fragmentation function. Note that for $\gamma$-h correlation $z_T \approx z$ since the trigger photon to leading order does not fragment (it may however be produced in a shower).

When comparing the results from A-A and p-p collisions, usually the yield ratios $I_{AA} = YPT_{med}/YPT_{vac}$ are formed where $I_{AA} = 1$ indicates the absence of a medium modification to the correlation strength. Unlike in the case of $R_{AA}$, there is no strong {\it a priori} expectation $I_{AA} < 1$ in the medium. Various biases (discussed below) are potentially capable of generating $I_{AA} >1$ for $R_{AA} < 1$ under the right kinematical conditions. Thus, $I_{AA}$ needs to be interpreted carefully.

\subsection{The perturbative hard process}

In LO pQCD, the production of two hard partons $k,l$ 
is described by
\begin{equation}
\label{E-2Parton}
  \frac{d\sigma^{AB\rightarrow kl +X}}{dp_T^2 dy_1 dy_2} \negthickspace 
  = \sum_{ij} x_1 f_{i/A}(x_1, Q^2) x_2 f_{j/B} (x_2,Q^2) 
    \frac{d\hat{\sigma}^{ij\rightarrow kl}}{d\hat{t}}
\end{equation}
where $A$ and $B$ stand for the colliding objects (protons or nuclei) and 
$y_{1(2)}$ is the rapidity of parton $k(l)$. The distribution function of 
a parton type $i$ in $A$ at a momentum fraction $x_1$ and a factorization 
scale $Q \sim p_T$ is $f_{i/A}(x_1, Q^2)$. The distribution functions are 
different for free protons \cite{CTEQ1,CTEQ2} and nucleons in nuclei 
\cite{NPDF,EKS98,EPS09}. The fractional momenta of the colliding partons $i$, 
$j$ are given by $ x_{1,2} = \frac{p_T}{\sqrt{s}} \left(\exp[\pm y_1] 
+ \exp[\pm y_2] \right)$.
Expressions for the pQCD subprocesses $\frac{d\hat{\sigma}^{ij\rightarrow kl}}{d\hat{t}}(\hat{s}, 
\hat{t},\hat{u})$ as a function of the parton Mandelstam variables $\hat{s}, \hat{t}$ and $\hat{u}$ 
can be found e.g. in \cite{pQCD-Xsec}.

To account for various effects, including higher order pQCD radiation, transverse motion of partons in the nucleon (nuclear) wave function and effectively also the fact that hadronization is not a collinear process, the distribution is commonly folded with an intrinsic transverse momentum $k_T$ with a Gaussian distribution, thus creating a momentum imbalance between the two partons as ${\bf p_{T_1}} + {\bf p_{T_2}} = {\bf k_T}$.

\subsection{The fragmentation process}

If the full distribution of hadrons in a jet generated from a parent parton is known, the distribution of leading and up to the $n$th order subleading hadrons in a jet can be written as an expansion of the particle distribution in the jet in terms of a tower of conditional probability densities $A_i(z_1, \dots, z_i, \mu^2)$ with the probability to produce $n$ hadrons with momentum fractions $z_1, \dots z_n$ from a parton with momentum $p_T\sim \mu$  being $\Pi_{i=1}^n A_i(z_1,\dots z_i,\mu^2)$ \cite{Dihadron_LHC}. 

Computing the full distributions of hadrons in a vacuum or medium-modified jet is in general a complicated task requiring a model for the non-perturbative hadronization in addition to the perturbative evolution of a parton shower. In the absence of a medium, Monte-Carlo (MC) codes such as PYTHIA \cite{PYTHIA} or HERWIG \cite{HERWIG} are used successfully for this purpose and can be used to compute the $A_i$. 

In the medium, the $A_i$ need to be computed from  a MC framework for in-medium shower evolution such as YaJEM \cite{YaJEM1,YaJEM2} or determined approximately from a vacuum shower code and a probability distribution  for leading parton energy loss $P(\Delta E)$.
In section \ref{sec:Elosses} we will describe how these objects are obtained from our models for the parton-medium interaction in detail. Combining the $A_i$ yields the full single-particle fractional momentum spectrum of produced hadrons, i.e. the medium-modified fragmentation function (MMFF) $D_{MM}^{f \rightarrow h}(z,\mu^2)$. Here $z$ is the fractional momentum of the hadron and $\mu$ the momentum scale. In the leading parton energy loss approxination, this quantity can also be computed as the convolution $D_{MM}^{f \rightarrow h}(z,\mu^2) = P(\Delta E) \otimes D^{f \rightarrow h}(z, \mu^2)$ where  $D^{f \rightarrow h}(z, \mu^2)$ is the vacuum fragmentation function.

As a side remark, while the expansion in terms of $A_i$ seems cumbersome when one has MC tools for the whole shower evolution available, it is nevertheless useful for two reasons. Conceptually it allows to understand the dependence of the subleading hadron distributions on the momentum of the leading shower hadron in a straightforward way since e.g. $A_2$ is a function of both $z_1$ and $z_2$. Technically, it allows to factorize the production of hard back-to-back parton pairs in the medium from the full in-medium shower evolution which makes the problem numerically easier to treat.

\subsection{Trigger bias in vacuum}

\begin{figure*}[htb]
\begin{center}
\epsfig{file=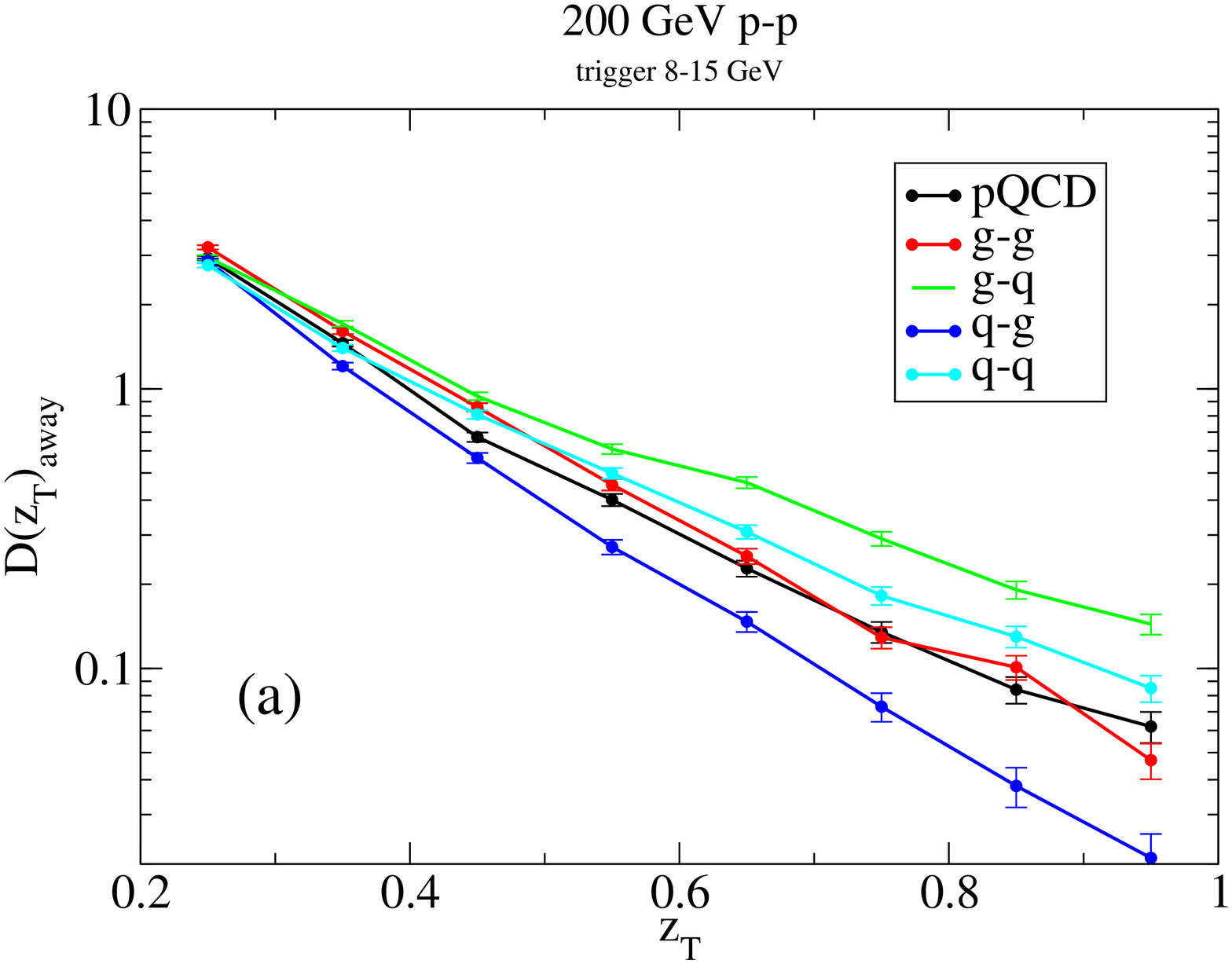, width=8.5cm}\epsfig{file=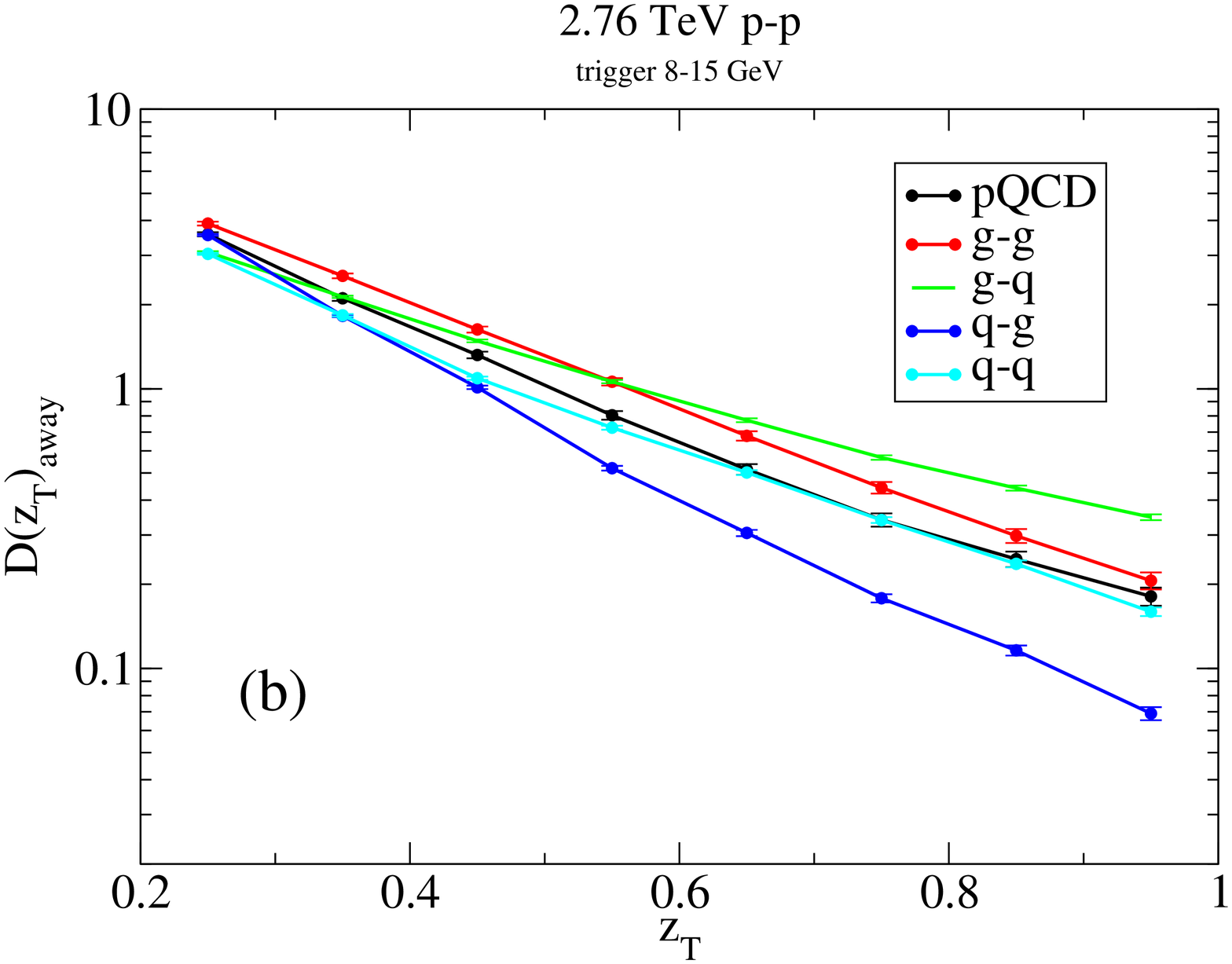, width=8.5cm}
\end{center}
\caption{\label{F-DzT-subproc}(Color online) The away side $z_T$ distribution of charged hadrons for RHIC kinematics (left panel) and LHC kinematics (right panel) in the absence of a medium, computed for the weight of subprocesses given by pQCD and under the assumption that only subprocesses with a particular out state is present. Here, the fragmentation is computed using HERWIG.}
\end{figure*} 

As mentioned above, triggering on hadrons in a certain momentum range is equivalent to selecting a subclass of events out of an unbiased distribution of back-to-back jets. In the absence of a medium, the trigger condition induces two distinct types of biases: a kinematic bias and a bias on parton type.

Let us define the energy of the back-to-back parton pair in its own restframe as $2E$. The momentum difference between pair restframe and c.m. frame of the p-p (A-A) collision is then $k_T$, leading to parton energies $E_1$ and $E_2$ in the c.m. frame. Assuming for simplicity for the parton rapidities $y_1 = y_2 = 0$ and negligible hadron and parton masses, the leading hadrons on near and away side are then found at $P_T = z_1 E_1$ (or $z_1 E_2$), subleading hadrons at $z_i E_1$ (or $z_i E_2$).  

The kinematic bias can be understood as follows: Since a trigger condition $P_T > P_{trig}$ for any given primary parton energy $E_1$ implies $z_1 > P_{trig}/E_1$, for sufficiently small $E_1$ of the order of $P_{trig}$ a hard fragmentation of the leading hadron $z_1 \sim 1$ is required. This in turn has implication for the distribution of subleading shower hadrons $A_2(z_1,z_2,\mu)$ since the phase space is restricted by the condition $\sum z_i = 1$. In other words, in this case the distribution of subleading hadrons correlated with the trigger softens as compared to the distribution in unbiased jets. The argument does not apply when $E$ is large, however the contribution from large parton energies are suppressed since the pQCD parton spectrum Eq.~(\ref{E-2Parton}) falls steeply $\sim 1/E^n$ with $n=7,8$ at RHIC kinematics and $n=4,5$ at LHC kinematics.

Another part of the kinematic bias concerns the momentum imbalance cast into the form of an intrinsic $k_T$. Since, in the extreme case where ${\bf k_T}$ and parton momentum are parallel, for the same fragmentation function a parton with $E_1 = E + k_T/2$ is more likely to meet the trigger condition than a parton with $E_2 = E - k_T/2$, the trigger condition biases the distribution towards situations in which $k_T$ is aligned with the trigger hadron momentum and away from situations in which $k_T$ points into the direction of the away side jet. The implication is that the momentum distribution on the away side is generically softer than on the near side. As in the previous case, the bias gradually disappears for $E \gg k_T$.

The bias on parton type is connected with the different fragmentation pattern of quarks and gluons --- quark jets exhibit a harder fragmentation pattern, i.e. are more likely to produce hard hadrons than gluon jets. A LO pQCD reaction has three possible out states: $gg, qq$ and $qg$. Imposing a trigger condition means thus that, for symmetric kinematical configurations, one is more likely to pick events in which a quark is in the final state than events with a gluon. In particular, for the $qg$ out state it also implies that typically the trigger hadron is part of the quark fragmentation whereas the away side jet corresponds to the gluon fragmentation. The importance of this latter effect depends on the $P_T$ dependent relative strength of the QCD subprocess $qg \rightarrow qg$.

In the following, we will distinguish between the out states $qg$ and $gq$. We take the first case as the situation where the trigger hadron is part of the quark fragmentation and the second case the situation where the trigger hadron was produced in the gluon fragmentation.

The combination of kinematic and parton type bias has a profound (and partially counter-intuitive) effect on the momentum distribution of away side hadrons. Consider a kinematical situation in which the away side momentum distribution is determined to a fraction $f$ by quark fragmentation (the larger $f$, the harder the distribution). Naively, in a gedankenexperiment, one would expect that when one hardens the quark fragmentation the away side momentum distribution to become harder as well. In a certain momentum window however, harder quark fragmentation will imply a higher $\langle z_1 \rangle$ for the trigger and therefore on average smaller partonic $E$ for the triggered events. However, for smaller $E$ the bias due to intrinsic $k_T$ becomes more important, which tends to soften the away side distribution. Likewise, if the $qg$ out state is a significant contribution competing with $gg$, then a harder quark fragmentation pattern will increase the fraction of quarks on the trigger side, implying through the $qg \rightarrow qg$ processes a lower fraction $f_1$ of quarks on the away side, in essence also softening the away side momentum distribution. Thus, under certain kinematical conditions, hardening the quark fragmentation may result in a counter-intuitive softening of the away side momentum distribution.

In Fig.~\ref{F-DzT-subproc} we illustrate the combination of kinematic and parton type bias by showing the result for the away side $z_T$ distribution for RHIC and LHC for an 8-15 GeV trigger range in the absence of a medium, computed with HERWIG \cite{HERWIG}, under the assumption that only those QCD subchannels which can produce the indicated partonic final state are active. For the flavour-equal out states $qq$ and $gg$, there is very little difference in slope, indicating that $D(z_T)$ does not reflect the fragmentation function $D(z,\mu^2)$ directly. However, for the flavour-unequal out states $qg$ and $gq$ substantial differences in slope are observed. At RHIC kinematics, the full pQCD result is chiefly an average between $qq$ and $qg$ which by accident happens to agree will with the $gg$ result.  At LHC, the biases are in general somewhat weaker, as the harder parton $p_T$ spectra permit to probe larger $E$ more easily. Here, the pQCD result is dominated by $gg$ and softened by some admixture of $qg$.

\subsection{Trigger bias in the medium}

When a back-to-back parton pair is embedded in a hot and dense medium, interactions with the medium generically soften the fragmentation pattern, and the effect increases with in-medium pathlength and medium density. This has implications for the kinematic and parton type bias and introduces a new geometrical bias.

Since the softening of the fragmentation increases with in-medium pathlength, the trigger hadron will more likely come from the parton which had the shorter path to the medium surface. In models, most partons leading to a trigger emerge from a region $\sim 4$ fm from the medium center in central collisions \cite{Dihadron}. This biases the away side parton to a long in-medium pathlength, and thus towards additional suppression, leading to the expectation $I_{AA}^{away} < 1$.

On the other hand, the softening of the fragmentation for the trigger implies $\langle z_1^{med} \rangle < \langle z_1^{vac} \rangle$ and therefore $\langle E^{med}\rangle > \langle E^{vac} \rangle$, i.e. the mean parton energy of triggered events is higher in the medium than in vacuum. Since this holds both for near and away side, for symmetric flavour configurations it leads to the expectation $I_{AA}^{away} > 1$ and counteracts the geometrical bias. At the same time, larger $\langle E^{vac} \rangle$ implies that any bias due to intrinsic $k_T$ is weaker than in vacuum, again leading to $I_{AA}^{away} > 1$.

Finally, a gluon interacts by a color factor of 9/4 more strongly with the medium than a quark. Thus, a medium acts like a filter for gluons and will enhance the fraction of quark jets leading to a trigger hadron relative to gluon jets. If the out state $qg$ is a significant contribution, this has implications for the away side which will then have a smaller fraction of quark jets than in vacuum, leading to the expectation $I_{AA}^{away} < 1$ since the gluon fragmentation is softer.

On the near side, smaller $\langle z_1^{med} \rangle$ implies more phase space for the production of subleading hadrons (since $\sum z_i = 1$ constrains the phase space) and hence $I_{AA}^{near} >1$. The role of the medium as a gluon filter also produces an effect into the same direction. However, the subleading partons in a shower are also explicitly subject to medium interactions which soften their distribution, arguing a trend $I_{AA}^{near} < 1$ at large $z_T$ .

In order to summarize and illustrate these effects in greater detail, let us present a simple example with semi-realistic numbers. Consider a hadronic trigger momentum of 10 GeV and assume for simplicity that we can neglect the effect of intrinsic $k_T$. For RHIC kinematics in vacuum, quarks are most likely to contribute to hard hadron production, and as a result $\langle z_1^{vac} \rangle \approx 0.7$, i.e. the typical parton momentum contributing to the 10 GeV trigger hadron yield is $\sim 14$ GeV. The presence of a medium softens the fragmentation, and as a result  $\langle z_1^{med} \rangle \approx 0.5$, thus for the same hadronic trigger momentum the typical parton momentum becomes 20 GeV due to the kinematic bias. At LHC kinematics, the dominant pQCD subchannel is $gg \rightarrow gg$ and thus the fragmentation pattern is softer even in vacuum. Combined with the harder primary spectrum of partons, one finds $\langle z_1^{vac} \rangle \approx 0.5$, implying a parton energy of 20 GeV (considerably more than at typical RHIC conditions). The presence of a medium further softens the fragmentation, leading to $\langle z_1^{med} \rangle \approx 0.3$ and a typical parton momentum of $\sim 33$ GeV given the hadron trigger.

The implication is that the away side before the interaction with the medium is hardened by the kinematic bias, and that the hardening is stronger at LHC than at RHIC. If one studies 8 GeV hadrons on the away side, one can again estimate some simple numbers using the parton energies estimated above. With the typical parton energies found above, one probes different regions of the fragmentation function, i.e. $z_1^{vac}\approx 0.57$ (RHIC vacuum), $z_1^{med/vac} \approx 0.4$ (RHIC medium, LHC vacuum) and  $z_1^{med} \approx 0.24$ (LHC medium). For a simple exponential ansatz for the fragmentation function $D(z) \sim \exp[-7z]$, this implies a factor $\sim 9$ more yield per trigger of hadrons at 8 GeV for LHC medium conditions than for RHIC vacuum conditions before the interaction with the medium on the away side, a striking illustration of the kinematic bias connecting near side trigger bias with the away side momentum spectrum. The geometrical bias to have on average a larger in-medium pathlength on the away side may still be strong enough to turn this kinematical enhancement into a net suppression, but this is not {\it a priori} obvious.

It is evident from these considerations that qualitative arguments are insufficient to interpret $I_{AA}$ with any degree of confidence, as the end result will be strongly influenced by a combination of biases acting into different directions. Any meaningful interpretation of experimental data requires thus a full simulation taking all effects into account with their expected magnitude.

\section{The framework}
\label{sec:Elosses}

The starting point for our computation of the high $P_T$ hadron yield in an A-A collision is Eq.~(\ref{E-2Parton}) which we evaluate at midrapidity $y_1 = y_2 = 0$. We sample this expression using a MC code introduced in \cite{Dihadron} by first generating the momentum scale of the pair and then the (momentum-dependent) identity of the partons. A randomly chosen $k_T$ with a Gaussian distribution of width 2.5 GeV is then added to the pair momentum. 

We assume that the distribution of vertices follows binary collision scaling as appropriate for a LO pQCD calculation. Thus, the probability density to find a vertex in the transverse plane is

\begin{equation}
\label{E-Profile}
P(x_0,y_0) = \frac{T_{A}({\bf r_0 + b/2}) T_A(\bf r_0 - b/2)}{T_{AA}({\bf b})},
\end{equation}
where the thickness function is given in terms of Woods-Saxon distributions of the the nuclear density
$\rho_{A}({\bf r},z)$ as $T_{A}({\bf r})=\int dz \rho_{A}({\bf r},z)$ and $T_{AA}({\bf b})$ is the standard nuclear overlap function $T_{AA}({\bf b}) = \int d^2 {\bf s}\, T_A({\bf s}) T_A({\bf s}-{\bf b})$ for impact parameter ${\bf b}$. We place the parton pair at a probabilistically sampled vertex $(x_0,y_0)$ sampled from this distribution with a random orientation $\phi$ with respect to the reaction plane.

We then propagate both partons on eikonal paths through the medium and compute for this path the leading and first two subleading fragments using the medium-modified conditional probability densities $A_1(z_1, \mu)$, $A_2(z_1, z_2, \mu)$ and $A_3(z_1, z_2, z_3, \mu) \approx A_2(z_1+z_2, z_3, \mu)$ given the path. Finally we check if there is a hadron in the event which fulfills the trigger condition. If not, we discard the event and start generating a new one. If there is a trigger hadron, we bin the remaining hadrons in the event on the near and away side in either $P_T^{assoc}$ or $z_T$. 

The key point in the computation are thus the various medium modified $A_i$, which in turn require knowledge of the fragmentation function. For the vacuum baseline, we extract the $A_i$ from either PYTHIA \cite{PYTHIA} or HERWIG \cite{HERWIG} (as will be seen below, HERWIG provides a slightly better description of the data, but since the in-medium shower code YaJEM utilizes the PYSHOW algorithm \cite{PYSHOW}, PYTHIA is the only consistent vacuum baseline choice if YaJEM is used for the medium modification). 

In the medium, the basic quantity characterizing the fragmentation pattern is the MMFF $D_{MM}(z, \mu^2,\zeta)$  given the parton path. Its computation requires knowledge of the geometry of the medium (e.g. in terms of a spacetime description of medium density) as well as a model for parton-medium interaction. Based on the results of \cite{JetHydroSys}, we choose a 2+1 d ideal hydrodynamical model \cite{hydro2d} (hydro I) and a 3+1d hydrodynamical model \cite{hydro3d} (hydro II) as extreme cases for the effect of the evolving medium geometry for RHIC kinematics. For LHC kinematics, we choose an extrapolation of the 2+1d model to $\sqrt{s} = 2.76$ ATeV using the EKRT saturation model as described in \cite{SpectraLHC}.

If the angle between outgoing parton and the reaction plane is $\phi$, the path of the parton through the medium $\zeta(\tau)$ (i.e. its trajectory $\zeta$ as a function of proper medium evolution time $\tau$) is determined in an eikonal approximation by its initial position ${\bf r_0}$  and the angle $\phi$ as $\zeta(\tau) = \left(x_0 + \tau \cos(\phi), y_0 + \tau \sin(\phi)\right)$ where the parton is assumed to move with the speed of light $c=1$ and the $x$-direction is chosen to be in the reaction plane. Based on line integrals along the trajectory, the MMFF (and hence the various $A_i$) can be computed directly from the in-medium shower code YaJEM \cite{YaJEM1,YaJEM2} as described below.

On the other hand, in the leading parton energy loss approximation the medium-modified perturbative production of hadrons can be computed from the convolution
\begin{equation}
D_{MM}(z, \mu^2,\zeta) \approx  P(\Delta E, \zeta) \otimes
D^{f \rightarrow h}(z, \mu^2)
\end{equation} 
where $P(\Delta E, \zeta)$ is the medium-induced energy loss probability given the path $\zeta$ and $D^{f \rightarrow h}(z, \mu^2)$ is the vacuum fragmentation function. The underlying assumption is that the dynamics of parton-medium interactions can largely be cast in terms of a shift in leading parton energy. In this case, the energy-loss probability for a given path $\zeta$ of a parton through the medium, $P(\Delta E, \zeta)$, is the ingredient to be computed within a specific model of parton-medium interaction.

The details of the parton-medium interaction model are thus in either the energy loss probability distribution $P(\Delta E,\zeta)$ for leading parton energy loss models or the MMFF $D_{MM}(z, \mu_p^2,\zeta)$ for in-medium shower models, given a specific path through the medium, and we will discuss how these are obtained in the next section.

In all cases, the nuclear modification factor is computed given the medium-modified yield of hard hadron production as
\begin{equation}
\label{E-RAA}
R_{AA}(p_T,y) = \frac{dN^h_{AA}/dp_Tdy }{T_{AA}({\bf b}) d\sigma^{pp}/dp_Tdy}.
\end{equation}
In the following we require for all computations that the parton-medium interaction model describes $R_{AA}$ in central collisions (i.e. the suppression of the rate of triggers in the medium) correctly both at RHIC and LHC and adjust model parameters accordingly. The results for the conditional yields are then obtained without additional free parameters.

\subsection{Armesto-Salgado-Wiedemann (ASW) formalism}

The detailed calculation of $P(\Delta E, \zeta)$ follows the Baier-Dokshitzer-Mueller-Peigne-Schiff (BDMPS) formalism for radiative energy loss  \cite{Jet2} using quenching weights as introduced by Salgado and Wiedemann \cite{QuenchingWeights}.

In this framework, the energy loss probability $P(\Delta E,\zeta)$ for a path can be obtained by evaluating the line integrals along $\zeta(\tau)$ as

\begin{equation}
\label{E-omega}
\omega_c({\bf r_0}, \phi) = \int_0^\infty \negthickspace d \zeta \zeta \hat{q}(\zeta) \quad  \text{and} \quad \langle\hat{q}L\rangle ({\bf r_0}, \phi) = \int_0^\infty \negthickspace d \zeta \hat{q}(\zeta)
\end{equation}
with the relation 

\begin{equation}
\label{E-qhat}
\hat{q}(\zeta) = K_{med} \cdot 2 \cdot \epsilon^{3/4}(\zeta) (\cosh \rho - \sinh \rho \cos\alpha)
\end{equation}
assumed between the local transport coefficient $\hat{q}(\zeta)$ (specifying the quenching power of the medium), the energy density $\epsilon$ and the local flow rapidity $\rho$ with angle $\alpha$ between flow and parton trajectory \cite{Flow1,Flow2}. $K_{med}$ is the adjustible parameter in this framework. It is naturally expected to be $O(1)$, but in fits to data at 200 AGeV the parameter takes (dependent on the precise hydrodynamical model) values ranging between 3 and 10 (the latter number occurs for viscous hydrodynamics where the initial entropy density is lower than in the ideal case, see \cite{JetHydroSys}).

Using the numerical results of \cite{QuenchingWeights} and the definitions above, the energy loss probability distribution given a parton trajectory can now be obtained as a function of the initial vertex and direction $({\bf r_0},\phi)$ as $P(\Delta E; \omega_c({\bf r},\phi), R({\bf r},\phi)) \equiv P(\Delta E,\zeta)$ for $\omega_c(\zeta)$ and $R=2\omega_c(\zeta)^2/\langle\hat{q}L(\zeta)\rangle$. 

\subsection{AdS/QCD hybrid model (AdS)}

The AdS/QCD hybrid model is a phenomenological model based on the ASW formalism which is described in \cite{StrongCoupling}.

Its basic assumption is that the hard scales in the process can be treated 
perturbatively as in the standard pQCD radiative energy loss calculations, 
while the interaction with the plasma which involves strong-coupling 
dynamics can be modeled {based on AdS/CFT considerations} for the $N{\,=\,}4$ 
super-Yang-Mills (SYM) theory \cite{StrongCoupling}. Instead of random transverse kicks, the strongly coupled medium exerts a force of magnitude $T^2$ (with $T$ the medium temperature) onto a parton and its  virtual gluon cloud, and in the AdS model this force rather than transverse broadening is responsible for placing virtual gluons on-shell and thus leading to induced radiation.
In a static medium of constant density, 
this {phenomenological} approach {suggests} an 
$L^3$ dependence of the mean energy loss {\cite{Gubser:2008as}}. 

While the model thus utilizes the same set of quenching weights as the ASW formalism described above, in order to reproduce a parametric $L^3$ dependence Eq.~(\ref{E-omega}) is replaced by

\begin{eqnarray}
\label{E-lineintegral3}
  \langle \hat{q} L \rangle({\bf r_0}, \phi) &=& K_{AdS} \int d\zeta\, \zeta\, T^4(\zeta),
\\
\label{E-lineintegral4}
  \omega_c({\bf r_0}, \phi) &=& K_{AdS} \int d\zeta\, \zeta^2\, T^4(\zeta),
\end{eqnarray}
with a different model parameter $K_{AdS}$.

\subsection{YaJEM (Yet another Jet Energy-loss Model)}

The MC code YaJEM is based on the PYSHOW code \cite{PYSHOW} which is part of PYTHIA \cite{PYTHIA}. It simulates the evolution from a highly virtual initial parton to a shower of partons at lower virtuality in the presence of a medium. A detailed description of the model can be found in \cite{YaJEM1,YaJEM2}.

The parton shower developing from a highly virtual initial hard parton in this model is described as a series of $1\rightarrow 2$ splittings $a \rightarrow bc$ where the virtuality scale decreases in each splitting, i.e. $Q_a > Q_b,Q_c$ and the energy is shared among the  partons $b,c$ as $E_b = z E_a$ and $E_c = (1-z) E_a$. The splitting probabilities for a parton $a$ in terms of $Q_a, E_a$ are calculable in pQCD and the resulting shower is computed event by event in a MC framework.  In the presence of a medium, the main assumption of YaJEM is that the parton kinematics or the splitting probability is modified. In the RAD (radiative energy loss) scenario, the relevant modification is a virtuality gain

\begin{equation}
\label{E-Qgain}
\Delta Q_a^2 = \int_{\tau_a^0}^{\tau_a^0 + \tau_a} d\zeta \hat{q}(\zeta)
\end{equation}
through the interaction with the medium. This modification leads to an increase in radiation. In order to evaluate Eq.~(\ref{E-Qgain}) during the shower evolution, the momentum space variables of the shower evolution equations need to be linked with a spacetime position in the medium. This is done via the uncertainty relation for the average formation time as

\begin{equation}
\label{E-Lifetime}
\langle \tau_b \rangle = \frac{E_b}{Q_b^2} - \frac{E_a}{Q_a^2}
\end{equation} 
and randomized splitting by splitting by sampling $\tau_b$ from the distribution

\begin{equation}
\label{E-RLifetime}
P(\tau_b) = \exp\left[- \frac{\tau_b}{\langle \tau_b \rangle}  \right].
\end{equation}

The evolution for any given parton in the shower evolution is terminated as soon as the parton reaches a minimum virtuality scale $Q_0$. The result of the partonic evolution in terms of a shower of low virtuality partons is then passed on to the Lund model \cite{Lund} to hadronize. The fractional longitudinal momentum distribution of the resulting hadron distribution corresponds to the MMFF of the various hadron species.

In principle, the full functional form of $\hat{q}(\zeta)$ could determine the MMFF, which would be computationally very expensive as a full MC simulation would be needed for every possible path in the medium. However, due to an approximate scaling law identified in \cite{YaJEM1}, it is sufficient to compute the line integral

\begin{equation}
\label{E-Qsq}
\Delta Q^2_{tot} = \int d \zeta \hat{q}(\zeta)
\end{equation}
in the medium to obtain the full MMFF  from a YaJEM simulation. The scaling law implies that the MC simulation has to be run only for a finite set of paths and makes a numerical solution of the geometry averaging possible.

As in the previous case, $K_{med}$ in Eq.~(\ref{E-qhat}) serves as the adjustable parameter of the model once $Q_0$ is chosen. YaJEM requires, dependent on the underlying hydrodynamical model, rather natural values for $K_{med}$ ranging from 0.6 to 2. 

\subsection{YaJEM-D}

In the default version of YaJEM, the minimum virtuality scale is fixed at $Q_0 = 0.7$ GeV. In the version YaJEM-D (dynamical computation of $Q_0$) \cite{YaJEM-Pathlength}, the formation length of the in-medium shower is forced to be within the medium length. This corresponds to the choice

\begin{equation}
\label{E-Q0}
Q_0 = \sqrt{E/L}
\end{equation}
which depends on both in-medium pathlength $L$ and shower-initiating parton energy $E$. This introduces an explicit pathlength dependence into the MMFF and improves the agreement with the experimentally observed split between in-plane and out of plane emission of high $P_T$ hadrons in non-central collisions \cite{PHENIX-RAA-RP} significantly. Together with the stronger pathlength dependence, YaJEM-D also predicts a strong rise of $R_{AA}$ with $P_T$ in angular averaged observables \cite{SpectraLHC} which agrees with what has been seen by ALICE \cite{ALICE-RAA}.

\section{Results}

\subsection{The away side baseline}

\begin{figure}[htb]
\begin{center}
\epsfig{file=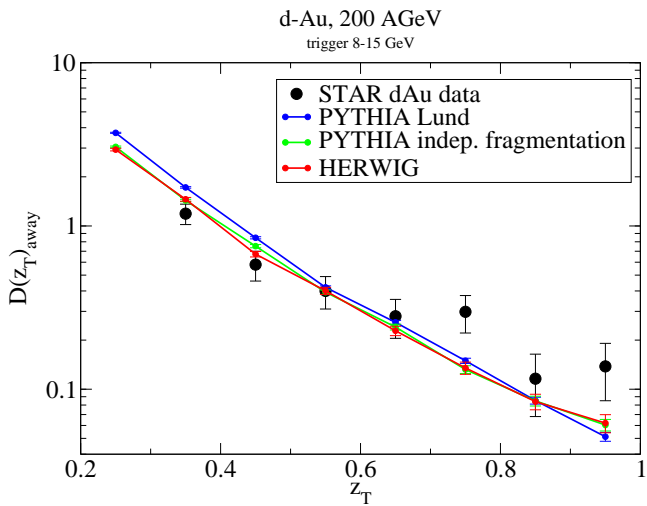, width=8.5cm}
\end{center}
\caption{\label{F-DzT-dAu}(Color online) The away side $z_T$ distribution for RHIC d-Au collisions, computed using PYTHIA with Lund hadronization, PYTHIA with independent fragmentation hadronization and HERWIG compared with the STAR data \cite{STAR-DzT}.}
\end{figure}

In Fig.~\ref{F-DzT-dAu} we show the obtained results compared with the baseline d-Au data in the absence of a medium for three different vacuum shower MC scenarios.

It is evident from the comparison with the data that all models exhibit a fragmentation pattern which is softer than what is seen in the data, with the default shower HERWIG being somewhat harder than the default PYTHIA with Lund model hadronization. This is not a novel observation: In the discussion of Fig.~1 in \cite{Dihadron_LHC} we have already pointed out that the fragmentation pattern of a default PYTHIA appears somewhat softer than a fragmentation function such as KKP which is based on a global fit to data. 

There is some evidence that the discrepancy with the data largely rests in the hadronization model underlying the shower codes, and could in principle be removed by suitable tuning of parameters --- if one replaces the Lund model by an independent fragmentation model also available in PYTHIA, the description of the data improves.

For the purpose of this paper, we will not attempt to tune the hadronization models to a best description of the data, but consider the output of HERWIG and PYTHIA with independent fragmentation as reasonably close to the data. Note also that the discrepancy is largely expected to cancel in a ratio like $I_{AA}$.

\subsection{The near side baseline}

\begin{figure}[htb]
\begin{center}
\epsfig{file=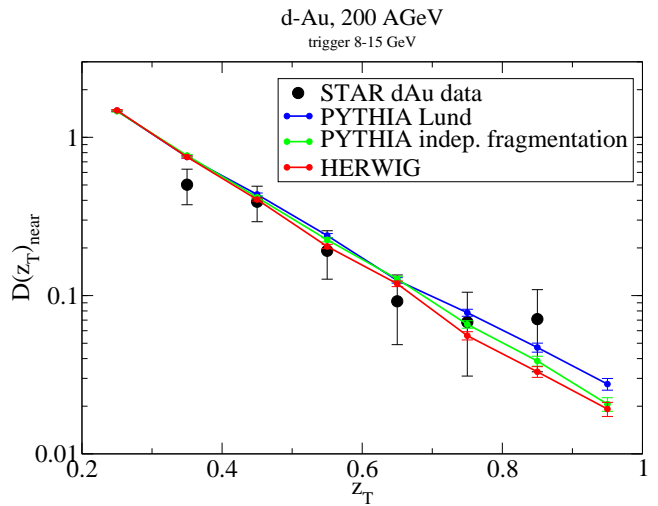, width=8.5cm}
\end{center}
\caption{\label{F-DzT-dAu-near}(Color online) The near side $z_T$ distribution for RHIC  d-Au collisions, computed using PYTHIA with Lund hadronization, PYTHIA with independent fragmentation hadronization and HERWIG compared with the STAR data \cite{STAR-DzT}.}
\end{figure}

In Fig.~\ref{F-DzT-dAu-near} we show the comparison of the vacuum shower evolution models with the d-Au data obtained by the STAR collaboration. In general, the d-Au data is described well by all three fragmentation functions, however the trend is reversed with respect to what was seen on the away side: Here PYTHIA with Lund hadronization (which has the softest fragmentation pattern) leads to the best agreement with the slope seen in the data. We note that this is in line with our previous discussion about the kinematic bias: If the fragmentation pattern is in general softer, the leading hadron will on average appear at lower $\langle z_1 \rangle$, thus leaving more phase space for the production of subleading hadrons. 

\subsection{The away side in a medium}

We proceed by embedding the simulation into the background of a hydrodynamically evolving medium. In order to test the sensitivity of the results to the detailed choice of the evolution model, we make use of the fact that for RHIC kinematics we with have two distinct hydrodynamical models available which in a previous study \cite{JetHydroSys} have been found to represent extreme cases in the underlying geometry, leading to the weakest (hydro I) and strongest (hydro II) dependence of high $P_T$ observables on medium geometry. The differences between the two hydrodynamical models are largely unconnected to the treatment of the longitudinal coordinate as scaling flow (hydro I) or full simulation (hydro II). Rather, high $P_T$ observables probe the space-time distribution of thermalized medium density, and hence properties such as the initialization time, the freeze-out conditions or the Equation of State being used.
Here, hydro I has an early equilibration time of $\tau_0 = 0.17$ fm and therefore also a large freeze-out temperature $T_F = 160$ MeV. In contrast, the equilibration time of hydro II is $\tau_0 = 0.6$ fm and the simulation ends at $T_F = 130$ MeV. Consequently, hydro II has a much larger freeze-out hypersurface as compared to hydro I, which is largely responsible for the observed modification of high $P_T$ probes.
For a more detailed comparison between the different scenarios, we refer the reader to \cite{JetHydroSys}.

\begin{figure*}[htb]
\begin{center}
\epsfig{file=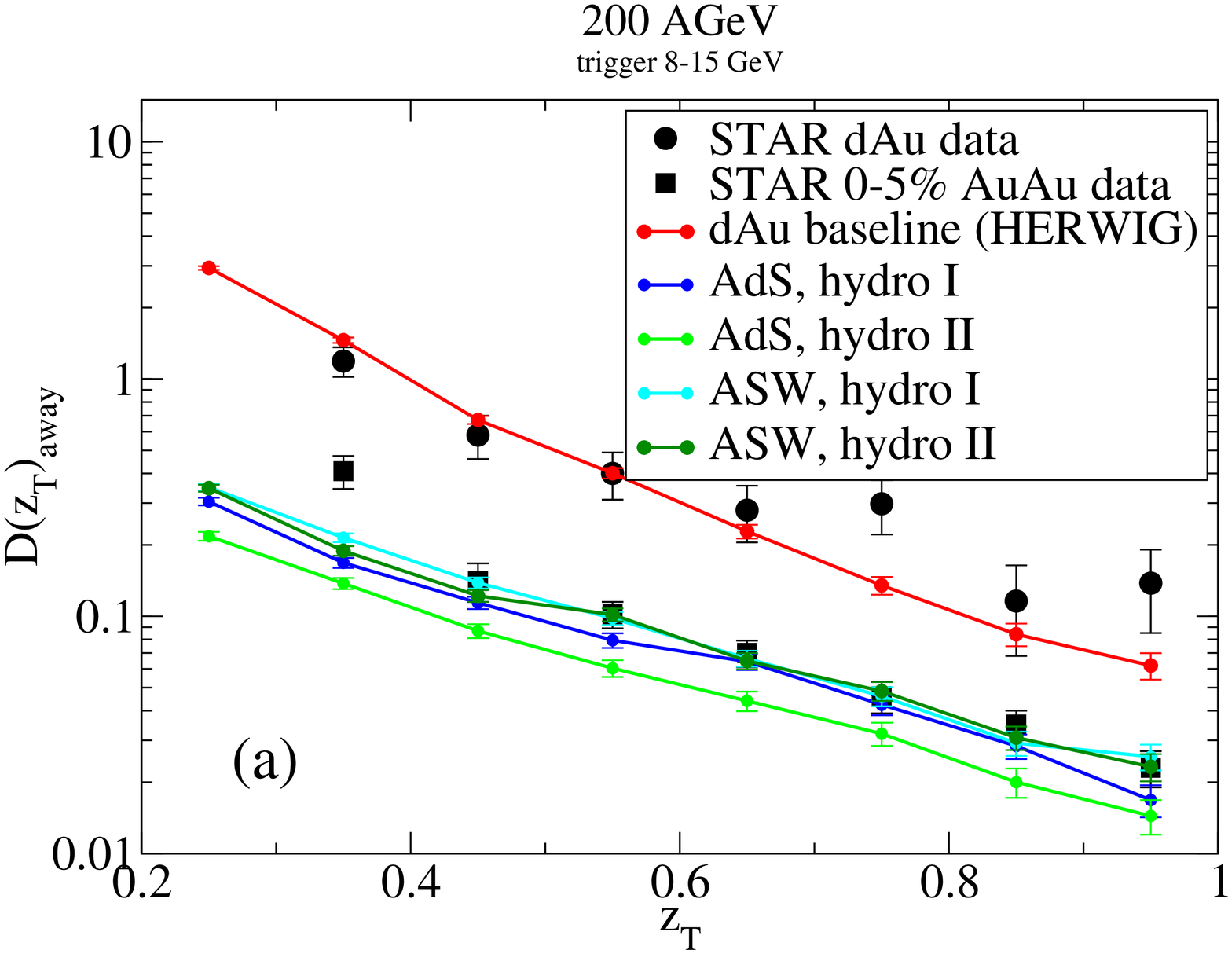, width=8.5cm}\epsfig{file=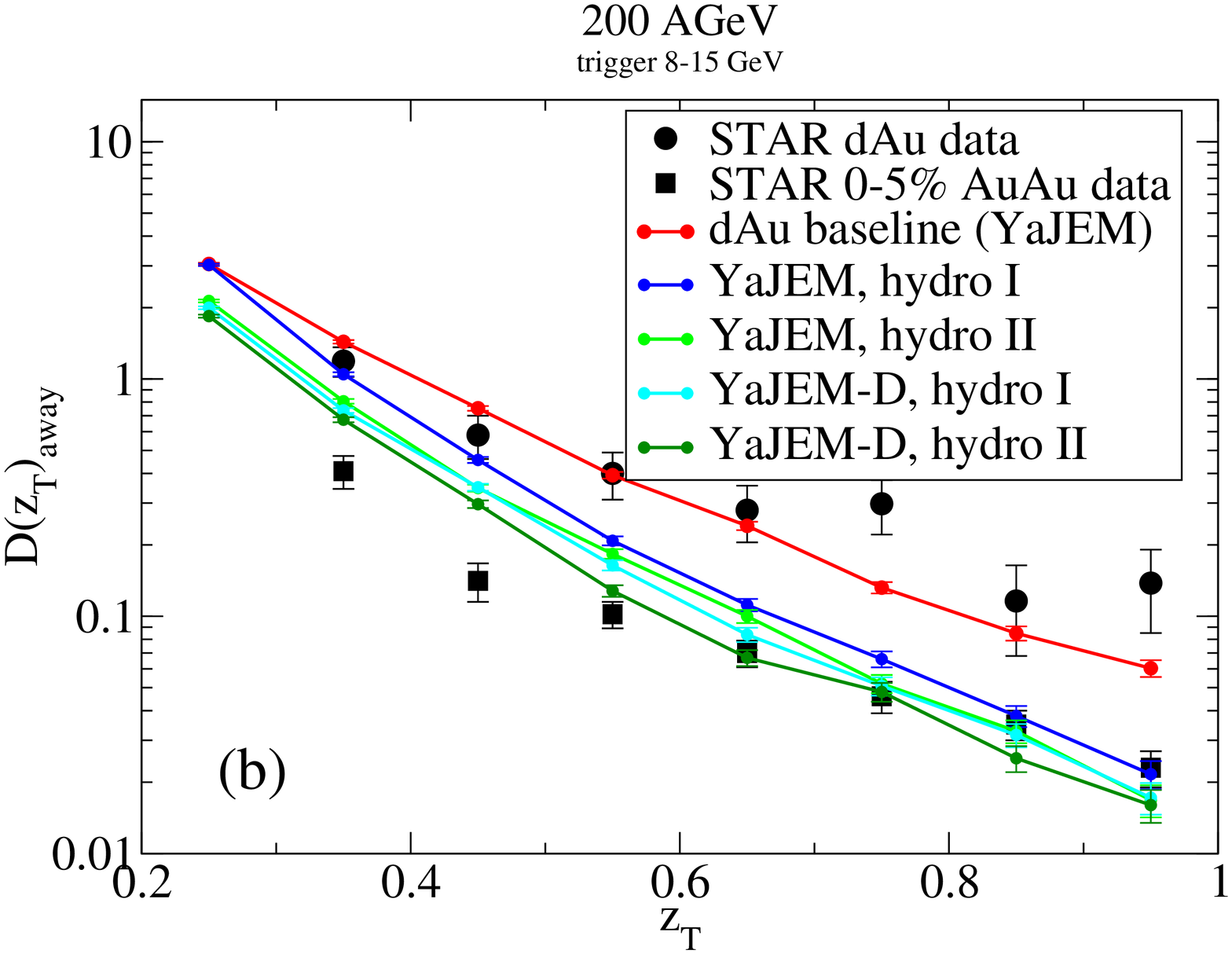, width=8.5cm}
\end{center}
\caption{\label{F-DzT-AuAu}(Color online) The away side $z_T$ distribution for RHIC kinematics for 200 AGeV dAu and 0-5\% central AuAu collisions, compared with various combinations of parton-medium interaction and bulk evolution models. Shown are leading parton energy loss models (left panel) and in-medium shower evolution models (right panel) for two different hydrodynamical scenarios (see text). }
\end{figure*}

In Fig.~\ref{F-DzT-AuAu} we show $D(z_T)$ for 0-5\% central Au-Au collisions at RHIC along with the appropriate baseline calculation, computed in all four models of parton-medium interaction using the two different hydrodynamical models for the bulk medium. As indicated earlier, the results for dihadron correlations in the medium are obtained without additional free parameters.

A few general trends can be observed: In all cases, the findings of \cite{JetHydroSys} are confirmed, i.e. the hydro II medium evolution leads to the most pronounced medium effect. The leading parton energy loss models tend to underpredict the data, whereas the in-medium shower evolution results show a trend to be above the data.

\begin{figure*}[htb]
\begin{center}
\epsfig{file=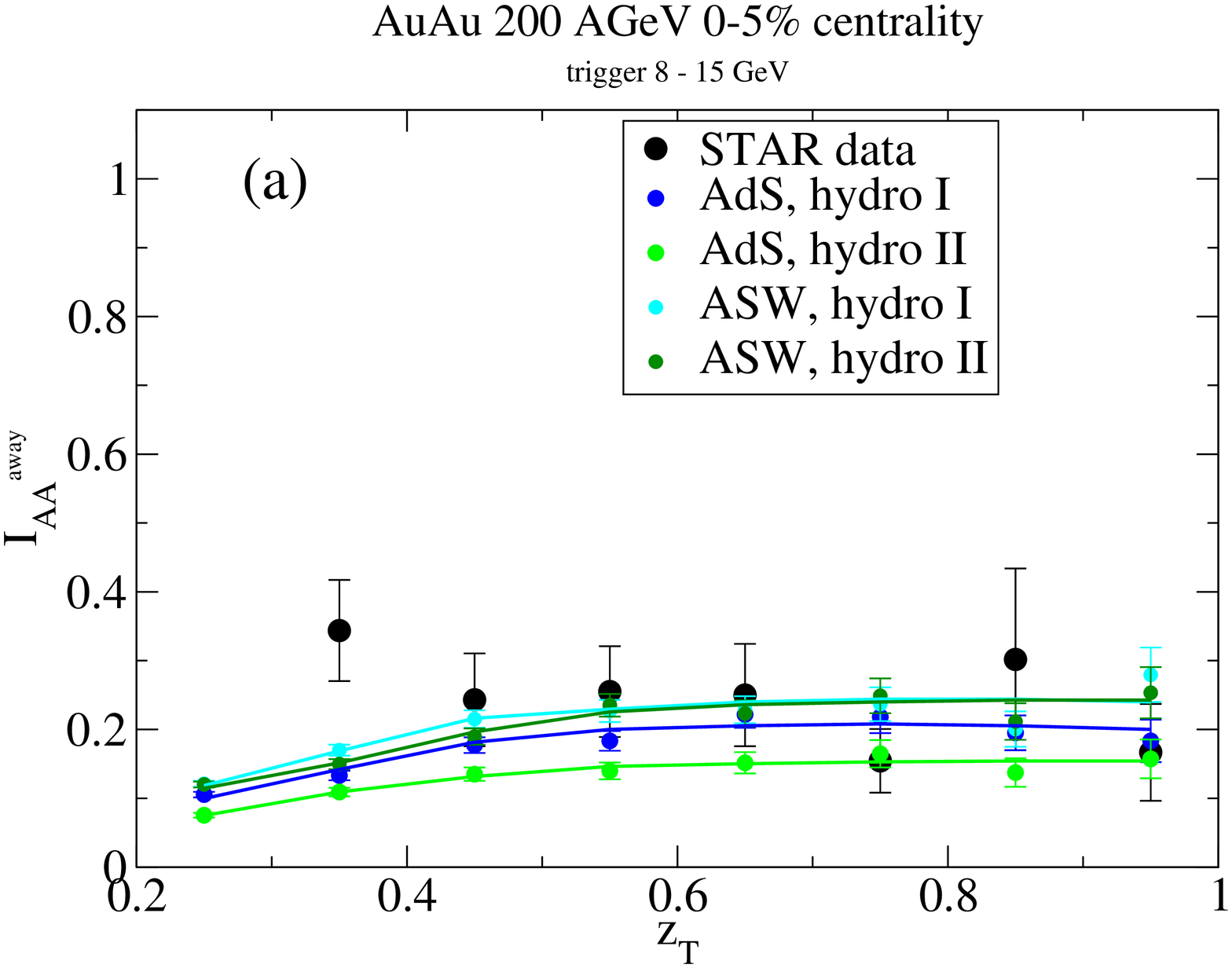, width=8.5cm}\epsfig{file=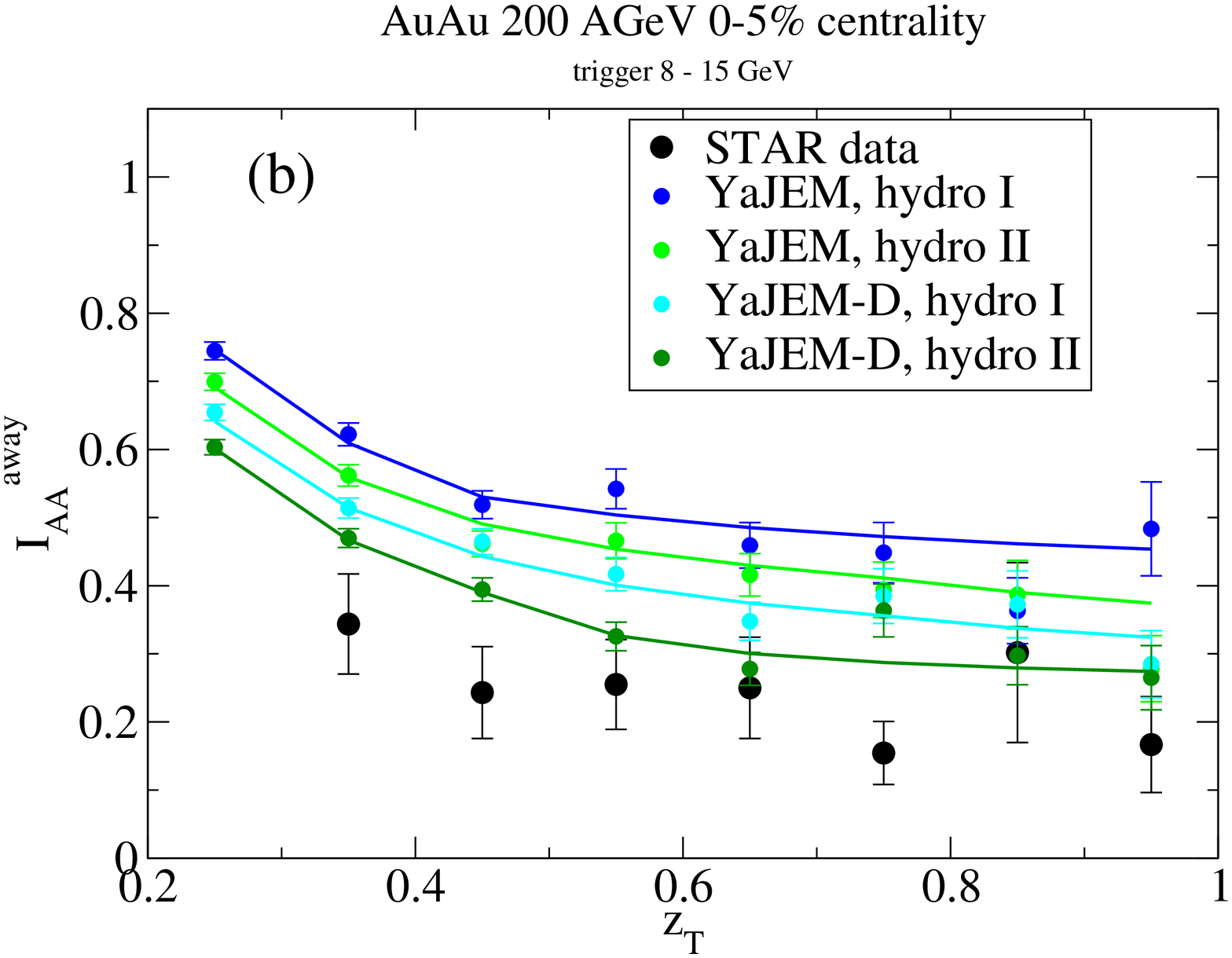, width=8.5cm}
\end{center}
\caption{\label{F-IAA-AuAu}(Color online) The away side $I_{AA}(z_T)$  for RHIC kinematics for 0-5\% central 200 AGeV AuAu collisions, compared with various combinations of parton-medium interaction and bulk evolution models. Shown are leading parton energy loss models (left panel) and in-medium shower evolution models (right panel) for two different hydrodynamical scenarios (see text). }
\end{figure*}

This is more clearly seen in Fig.~\ref{F-IAA-AuAu} where we form the ratio with the baseline calculation and compare with the data in terms of $I_{AA}(z_T)$. The following pattern emerges: In addition to the dependence on the medium model $I_{AA}^{hydro II} < I_{AA}^{hydro I}$ noted above, also $I_{AA}^{AdS} < I_{AA}^{ASW} < I_{AA}^{YaJEM-D} < I_{AA}^{YaJEM}$ is observed. This ordering reflects the pathlength dependence of the parton-medium interaction models, i.e. is caused by the geometrical bias, and agrees with the results obtained in \cite{JetHydroSys} and \cite{YaJEM-Pathlength}. Not all combinations of medium evolution and parton-medium interaction model agree with the data --- in particular AdS with hydro II underpredicts the data whereas YaJEM-D with the hydro I overpredicts the data. In line with the results of \cite{YaJEM-Pathlength}, the pathlength dependence of YaJEM is too weak to lead to sizeable geometrical bias and hence YaJEM fails to describe the data regardless of the assumed medium evolution model. In contrast, the ASW model describes the data rather well with both hydrodynamical models (ASW in combination with hydro I corresponds to one of the scenarios presented in our previous work \cite{Dihadron}). 

Let us next turn to details of the $z_T$ dependence: Here, the leading particle energy loss models show a qualitatively wrong behaviour at low $z_T$ --- while the data turn upward, the model results show the opposite trend. Given that the region $z_T < 0.5$ probes the way subleading hadrons in the shower are affected by the medium, it should not be a surprise that the leading parton energy loss approximation is insufficient for this region (as indeed was already observed in \cite{Dihadron}). The full in-medium shower codes on the other hand show the qualitatively correct upward turn at low $z_T$, however they predict an onset at higher $z_T$ than seen in the data. This is a non-trivial constraint for the in-medium shower modelling, and indicates that the mechanism to transport energy towards low $z$ in the shower is more efficient in nature than assumed by YaJEM. Presumably, an additional small component of elastic energy loss is capable of improving the description in the low $z_T$ region. Note that a similar failure of YaJEM has also been observed in $\gamma$-h correlations \cite{gamma-h}.

\begin{figure}[htb]
\begin{center}
\epsfig{file=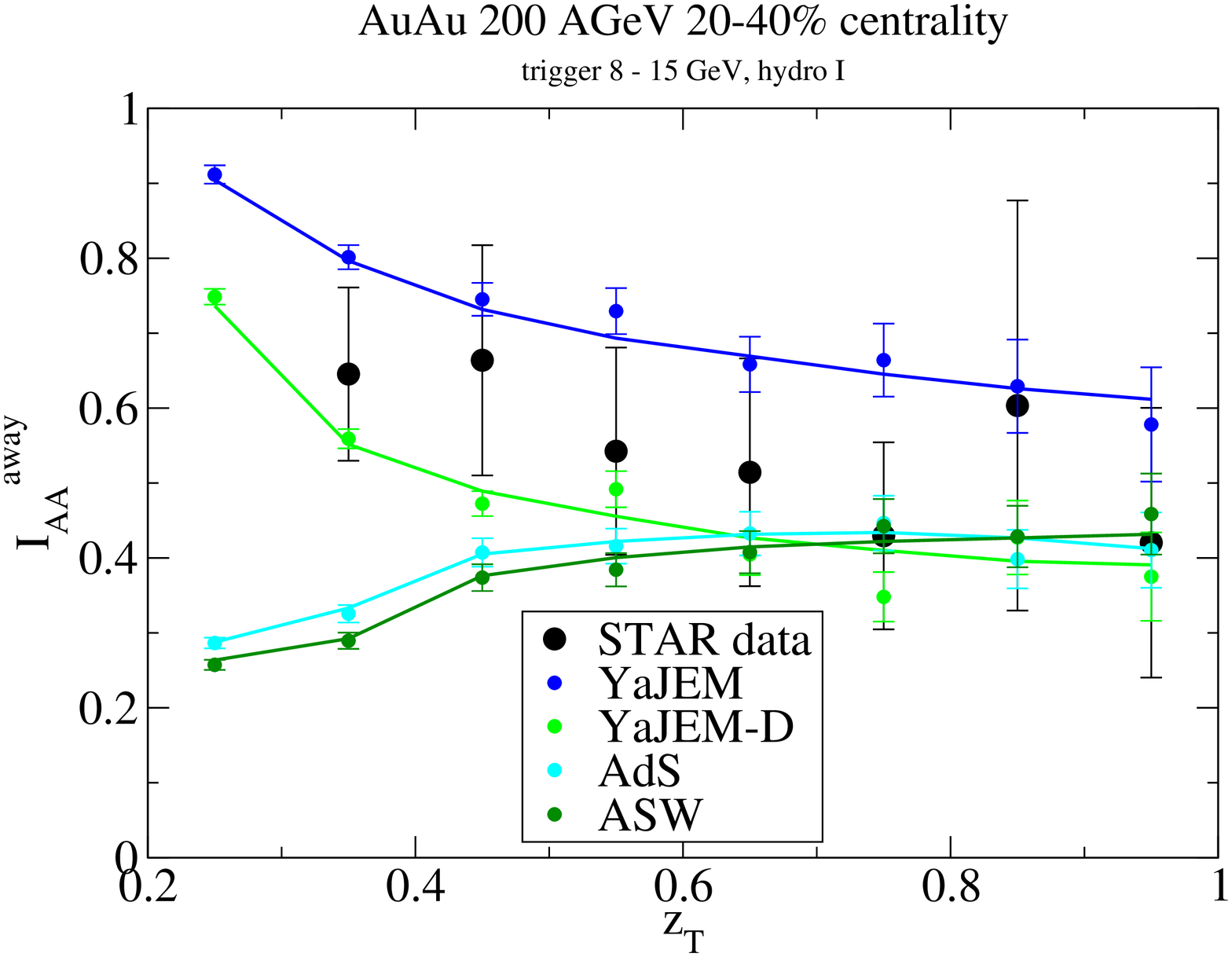, width=8.5cm}
\end{center}
\caption{\label{F-IAA-AuAu20-40}(Color online) The away side $I_{AA}(z_T)$ for 200 AGeV Au-Au collisions with 20-40\% centrality, computed using a 2+1d hydrodynamical medium description combined with various models for the parton-medium interaction (see text) compared with the STAR data \cite{STAR-DzT}.}
\end{figure} 

In order to probe a different density and pathlength distribution, we also discuss 20-40\% central collisions in Fig.~\ref{F-IAA-AuAu20-40} (for which we only have results from the 2+1d hydrodynamical evolution available). Here, YaJEM-D shows the best agreement with the data. The fact that YaJEM-D here slightly undershoots the data whereas for central collisions with the same 2+1d evolution model the results were above the data illustrates the strong non-linear pathlength dependence of the model.  Clearly, there is currently no combination of medium evolution and parton-medium interaction model which gives a completely satisfactory description of all the available data, thus future work on the details is needed.

\begin{figure}[htb]
\begin{center}
\epsfig{file=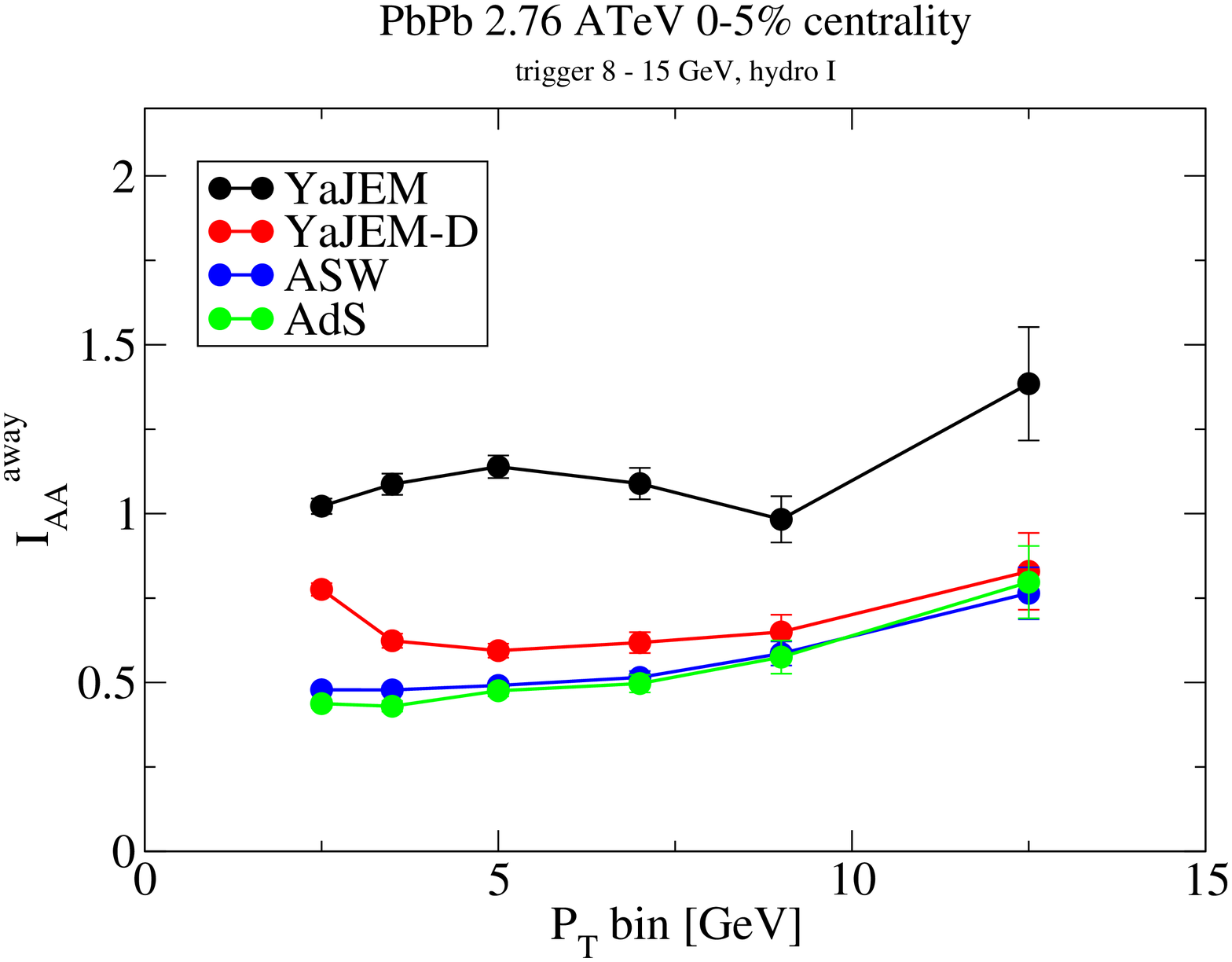, width=8.5cm}
\end{center}
\caption{\label{F-IAA-LHC_away}(Color online) The away side $I_{AA}(P_T)$ for 2.76 AGeV Pb-Pb collisions in the 0-5\% centrality class, computed using a 2+1d hydrodynamical medium description combined with various models for the parton-medium interaction (see text).}
\end{figure} 

Finally, we conclude the discussion of the medium-modified away side by presenting $I_{AA}(P_T)$ (the choice of $P_T$ rather than $z_T$ is motivated by the analysis strategy of the ALICE collaboration) for central 2.76 ATeV Pb-Pb collisions in Fig.~\ref{F-IAA-LHC_away}. This complements our discussion of the extrapolation of the nuclear suppression factor $R_{AA}$ from RHIC to LHC done in \cite{SpectraLHC}. Also for this computation, for the sake of clarity no retuning of parameters is done, although in the case of intrinsic $k_T$ a good case for a higher value can be made. Note that we only have results from a 2+1d hydrodynamical calculation \cite{SpectraLHC} (the extrapolation of hydro I to LHC energies) available at this point. It is readily apparent that $I_{AA}^{LHC} > I_{AA}^{RHIC}$. While this is surprising if one thinks in terms of geometrical bias only, the finding is readily understood in terms of the combination of kinematical and parton type bias. While for the 8-15 GeV trigger range at RHIC the $qg$ out state dominates, at the LHC almost exclusively $gg$ events contribute. Both the lower $\langle z_1 \rangle$ of the trigger hadron and the absence of a strong asymmetric contribution from $qg$ events tend to harden the away side momentum distribution with respect to RHIC conditions. 

The remaining results can be understood in terms of the geometrical bias probing the pathlength dependence of the parton-medium interaction model. For a weak pathlength dependence such as found in YaJEM, the kinematical bias dominates and $I_{AA} \approx 1$. For stronger pathlength dependence there is some noticeable geometrical bias, leading to $I_{AA} \sim 0.5$ with no substantial differences between YaJEM-D, ASW or AdS. 

\subsection{The near side in a medium}

It should be noted for the discussion of the near side correlation that unlike the away side, the dominant contribution to the high $z_T$ part of the near side correlation is given by the first subleading shower hadron. Thus, while the baseline calculation, being based on shower codes, is expected to contain all relevant physics, this is no longer true for the in-medium result obtained within the leading parton energy loss approximation.

No significant modification of the near side associate hadron momentum distribution has been observed or expected for RHIC kinematics \cite{Dihadron}, and this remains true for any of the models investigated here here. However, a near side $I_{AA} > 1$ was predicted in \cite{Dihadron_LHC}, albeit for a very different trigger range, and traced back to the parton type bias.

\begin{figure}[htb]
\begin{center}
\epsfig{file=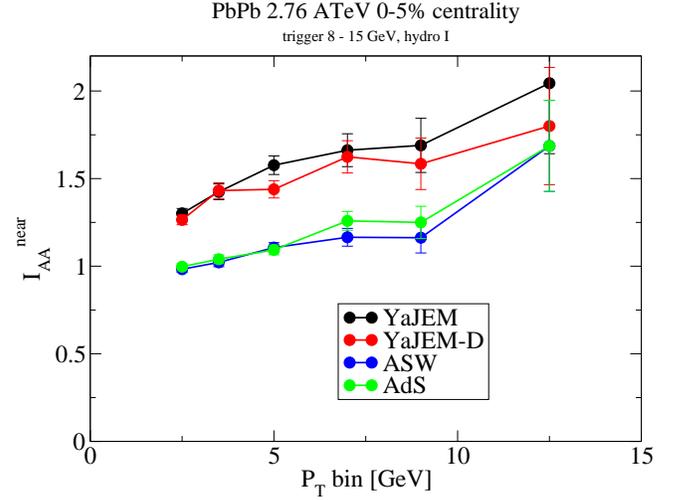, width=8.5cm}
\end{center}
\caption{\label{F-IAA-LHC_near}(Color online) The near side $I_{AA}(P_T)$ for 2.76 AGeV Pb-Pb collisions in the 0-5\% centrality class, computed using a 2+1d hydrodynamical medium description combined with various models for the parton-medium interaction (see text).}
\end{figure} 

In Fig.~\ref{F-IAA-LHC_near} we show the results for the near side associate hadron momentum yield (i.e. the modification of the subleading shower hadron momentum distribution) for 0-5\% central Pb-Pb collisions at 2.76 ATeV. All of the parton-medium interaction models predict an enhancement of $I_{AA}$ above unity for the 8-15 GeV trigger range, however the expected enhancement is much stronger in the in-medium shower models than in the leading parton energy loss models. Given that in-medium shower models explicitly include the dynamics of subleading shower parton interaction with the medium whereas energy loss models only influence this distribution via the kinematic and parton type bias, this finding is perhaps not a surprise.

The observed enhancement is strongest at high $P_T$. Note that events in this bin require a substantial amount of energy, i.e. in addition to a subleading hadron above 10 GeV also a triggered hadron with even larger energy and the energy lost from the medium. Consequently, the kinematic bias forces the parton energy to be very high, thus probing a region where the pQCD parton spectrum flattens and where ample phase space for secondary production is available.

\section{Discussion}

Let us reiterate the main results of the previous section in order to assess the potential of hard dihadron correlations as an observable. 

\begin{itemize}
\item The momentum spectrum of the away side correlated yield  is sensitive to both medium geometry and to the pathlength dependence of the parton-medium interaction model. While this is also true for single inclusive hadron observables such as $R_{AA}(\phi)$, the nuclear suppression factor as a function of the angle $\phi$ with respect to the bulk matter event plane \cite{JetHydroSys}, the pathlength distribution in the background geometry is very different. Thus, a combined analysis of $R_{AA}(\phi)$ and $I_{AA}(\phi)$ for different collision centralities might help to disentangle the combined effects of geometry in the hydrodynamical model and pathlength dependence inherent in the parton-medium interaction model.

\item The shape of the away side $z_T$ (or $P_T$) distribution probes energy redistribution in the shower, and is hence able to probe beyond leading particle energy loss. In the present investigation, we found constraints for YaJEM which were not available from single hadron observables. It should however be noted that this set of constraints might be identical with the ones which can be obtained from $\gamma$-h correlations \cite{gamma-h}.

\item The near side distribution likewise probes energy redistribution in the shower, but it does so with a kinematic bias in the fragmentation pattern to observe the trigger hadron. Thus this information is more differential than what can be found on the away side.

\item In certain kinematical regions, the parton type bias can be exploited to study a gluon-rich sample of jets. This complements the idea to study $\gamma$-h correlations in order to probe a quark-rich sample of jets, since $qg \rightarrow q\gamma$ dominates over the annihilation $q \overline{q} \rightarrow g\gamma$.

\end{itemize}

Thus, our results suggest that hard dihadron correlations are indeed a suitable tool to probe into the development of parton showers in the medium in a controlled way, starting from the leading hadron to the next subleading hadrons, while allowing for a clean separation between jet and medium physics at all times.

In the present investigation, only in-medium shower evolution models capture correctly the qualitative behaviour of the away side $z_T$ distribution seen in the data. Among the two in-medium shower evolution models we tested, only YaJEM-D remains a viable candidate since the pathlength dependence of the medium modification is too weak for YaJEM, and thus the away side $I_{AA}$ overpredicts the data by a large margin. However, even YaJEM-D is only viable with the large spatially extended medium predicted in the 3+1d hydrodynamical model. Thus, the hard dihadron correlations provide strong constraints both for the in-medium shower models and for hydrodynamical modelling.

\begin{acknowledgments}
 
We thank Steffen Bass, Chiho Nonaka and Hannu Holopainen for providing the hydrodynamical evolutions used in this work. T.R. is supported by the Academy researcher program of the
Academy of Finland, Project No. 130472. Additional support comes
from KJE's Academy Project 133005.
 
\end{acknowledgments}

\end{document}